\documentclass[12pt]{article}
\usepackage[section]{placeins}
\usepackage{float}
\usepackage{times}
\usepackage{amsmath}
\usepackage[round]{natbib}
\usepackage{epsfig}
\usepackage{amsfonts}
\usepackage{graphicx,subfigure,url}
\usepackage{authblk}
\usepackage{epsfig}
\usepackage{bm}
\usepackage{epstopdf}
\usepackage{graphicx}
\usepackage[a4paper]{geometry}
\usepackage{booktabs,multirow,tabularx}
\usepackage{fancybox}
\usepackage{pstricks}
\newcommand\nmfootnote[1]{%
  \begingroup
  \renewcommand\thefootnote{}\footnote{#1}%
  \addtocounter{footnote}{-1}%
  \endgroup
} 

\newcommand{\be}{\begin{eqnarray}}         
\newcommand{\ee}{\end{eqnarray}}           \newcommand{\ba}{\begin{eqnarray*}}
\newcommand{\ea}{\end{eqnarray*}}

\newcommand{\keywords}[1]{\par\noindent{\small{\em keywords\/}: #1}}

\begin{document}
 \title{WavmatND: A {MATLAB} Package for  Non-Decimated Wavelet Transform and its Applications }

\author{Minkyoung Kang}
\author{Brani Vidakovic}
\affil{{\small \emph{Georgia Institute of Technology, Atlanta, GA}}}
\date{}
\maketitle
\begin{abstract}
A non-decimated wavelet transform (NDWT) is a popular version of wavelet transforms because of its many advantages in applications. The inherent redundancy of this transform proved beneficial in tasks of signal denoising and
scaling assessment. To facilitate the use of NDWT, we  built a {MATLAB}
 package, {\bf WavmatND}, which has three novel features:  First, for signals of moderate size the proposed method reduces computation time of the NDWT by replacing repetitive convolutions with matrix multiplications. Second, submatrices of an NDWT matrix can be rescaled, which enables a straightforward inverse transform. Finally, the method has no constraints on a size of the input signal in one or in two dimensions, so signals of non-dyadic length and rectangular two-dimensional signals with non-dyadic sides can be readily transformed. We provide illustrative examples  and a tutorial to assist users in application of this stand-alone package.
 \end{abstract}
  \\ \\
 Submitted to {\it Journal of Statistical Software}
\section{Introduction}
In the last two decades, a number of applications in signal processing, such as data compression, signal de-noising, scaling assessment, and image processing, have benefited from advances in wavelet-defined multiscale methodologies. A wavelet transform reveals information hidden in the domain of data acquisition by looking at the interplay of time/scale properties in the transformed data. Signals can be transformed into time/scale domain by wavelets in many ways. Each version of a wavelet transform has characteristics that are useful in certain applications. A popular version is a non-decimated wavelet transform (NDWT), which overcomes some shortcomings of the standard orthogonal wavelet transform. The NDWT is a redundant transform  because it is performed by repeated filtering with a minimal shift
(or with a maximal sampling rate) at all dyadic scales. Subsequently, the transformed signal contains the same number of coefficients  as the original signal  at each multiresolution level. Because the NDWT does not decimate wavelet coefficients, the size of a transformed signal increases by its original size with each added decomposition level, and thus, the NDWT is computationally more expensive.  The efficiency of computation becomes particularly important for multi-dimensional signals. \\
In this paper, we describe a { MATLAB}$^\copyright$\nmfootnote{
${}^\copyright$ 2016 The MathWorks, Inc. MATLAB and Simulink are registered trademarks of The MathWorks, Inc.
See \url{www.mathworks.com/trademarks} for a list of additional trademarks.} package, {\bf WavmatND}, which efficiently performs the NDWT. The proposed package has three novel features. The first feature is that instead of using convolution-based Mallat's pyramid algorithm \cite{mallat1989theory}, we perform the NDWT by matrix multiplication. The matrix is formed directly from wavelet filter coefficients. \cite{remenyi2014image} also performed the NDWT using a matrix-based approach; however, their rules of constructing a matrix were based on Mallat's algorithm. \cite{percival2006wavelet} provide a matrix construction rule for NDWT, but the construction requires a convolution of filters in defining entries of the matrix, which is, essentially, Mallat's algorithm. The proposed method explicitly defines each entry of the transform matrix directly from the filter elements. With its simple construction rules, the proposed matrix-based NDWT requires significantly less time for computation compared to the convolution-based NDWT when the input signals are of a moderate size. \\

The second feature is that inverse transform matrix differs from the
transpose of direct transform matrix up to a multiplicative rescaling matrix.
Rescaling of  submatrices of a NDWT matrix is needed to both obtain resulting wavelet coefficients in their proper scales and retrieve the original signal without loss of information. Unlike the matrix for the orthogonal wavelet transform, which is a square matrix, a NDWT matrix for a $p$-depth decomposition of a signal of size $m$ consists of $(p+1)$ square ($[m \times m]$) submatrices, each of which corresponds to one decomposition level.
For a perfect reconstruction, the proposed process utilizes a weight matrix of size $[(p+1)\cdot m \times (p+1)\cdot m]$ that enables lossless reconstruction. The multiplication of the transposed NDWT matrix, the weight matrix, and the NDWT matrix, in that order, yields an identity matrix of size $[ m \times m ]$, which guarantees a  lossless inverse transform.
The matrix of \cite{percival2006wavelet} can retrieve an input signal but the resulting wavelet coefficients are down-scaled because of insisting on the energy preservation in redundant transform. With the proposed two-stage process, we can obtain the wavelet coefficients in their correct scale and then we can utilize a weight matrix if the inverse transform is necessary.\\

The third feature is that the package can handle one- or two-dimensional (1-D or 2-D) signals of an arbitrary size, and even the rectangular shapes in the case of a 2-D transform. This property is not shared by critically sampled wavelet transforms that require an input of dyadic size. In addition, one can perform a 2-D NDWT with two different wavelet bases, one base acting on the rows and another acting on the columns of the 2-D input signal, which allows for more modeling freedom in the case of spatially anisotropic 2-D signals.

The rest of the paper is organized into five sections. Section 2 introduces background information related to the NDWT, and Section 3 discusses the advantages of the matrix-based NDWT with simulation results. Section 4 illustrates the application of the package, while Section 5 provides a tutorial for using the package with simple example codes. Section 6 provides concluding remarks.

\section{Non-decimated Wavelet Transforms}
Unique characteristics of the NDWT are well captured by its alternative names such as ``stationary wavelet transform,'' ``time-invariant wavelet transform,'' ``{\it \'a trous} transform,'' or ``maximal overlap wavelet transform.'' In this section, we will overview the features of the NDWT that motivate  such names, beginning with a description of a one-dimensional NDWT for a discrete input.

Assume that a multiresolution framework is specified and that $\phi$ and $\psi$ are scaling and wavelet
functions respectively.
We represent a data vector ${\bm y}=(y_0,y_1, \dots, y_{m-1})$ of size $m$ as a function $f$  in terms of shifts of the scaling
function at some multiresolution level $J$ such that  $J-1 < \log_2 m \leq J$, as
\ba
f(x)= \sum_{k=0}^{m-1} y_k \phi_{J,k}(x),
\ea
where $\phi_{J,k}(x) = 2^{J/2} \phi\left(2^J (x - k)\right).$
The data interpolating function $f$ can be re-expressed as
\be
\label{eq:ndseries}
f(x)= \sum_{k=0}^{m-1} c_{J_0,k} \phi_{J_0,k}(x) + \sum_{j=J_0}^{J-1} \sum_{k=0}^{2^n-1}  d_{jk}2^{j/2}\psi\left(2^j( x -k )\right),
\ee
where
\ba
& &\phi_{J_0,k}(x) = 2^{J_0/2} \phi\left( 2^{J_0} (x - k) \right), \\
& &\psi_{jk}(x) = 2^{j/2}  \psi\left(2^{j} (x  - k) \right), \\
& & ~~~~~~~~j = J_0, \dots, J-1;~ k=0,1,\dots, m-1.
\ea

The coefficients, $c_{J_0, k}, k=0,\dots m-1$ and $d_{jk}, j=J_0,\dots,J-1; k=0, \dots, m-1,$
comprise the NDWT of vector $\bm y$.

 Notice that a  shift, $k$, is constant at all levels, unlike the traditional orthogonal wavelet transform in which the
 shifts are level dependent, $2^{-j}k$. This constancy of the shifts across the levels in NDWT
  indicates that the transform is time invariant. As we see from equation (\ref{eq:ndseries}), the NDWT produces a redundant representation of the data. For an original signal of size $m$ transformed into $p$ decomposition levels (the depth of transform is $p$), the resulting non-decimated wavelet coefficients are
  $\bm c^{(J_0)} = \left(c_{J_0,0},\dots, c_{J_0, m-1}\right)$ and
  ${\bm d}^{(j)} = \left(d_{j,0},\dots, d_{j, m-1}\right), ~j=J_0,\dots,J-1$, for $p= J-J_0.$
  Since NDWT does not decimate, nothing stops the user from taking $p$ larger than $\lceil \log_2 m \rceil.$ For such $p$ coarse levels of detail become zero-vectors.

   Coefficients in ${\bm d}^{(j)}$ serve as the detail coefficients while coefficients in $\bm c^{(J_0)}$ serve as the coarsest approximation of the data. Later, we will refer to these coefficients as d-type and c-type coefficients.
   With $p$ detail levels, the total number of wavelet coefficients is  $(p+1) \times m$.
   Such wavelet coefficients at different decomposition levels are related to one another by Mallat's pyramid algorithm (\cite{mallat1989multiresolution},   \cite{mallat1989theory}) in which convolutions of low- and high-pass wavelet filters, ($h$) and ($g$), respectively, take place in a cascade. The filters $h$ and $g$ are known as quadrature mirror filters. Given a low-pass wavelet filter $h=(h_0, \dots, h_M),$ fully and uniquely specified by the choice of wavelet basis, the $i^{th}$ entry of the high-pass counterpart $g$ is $g_i= (-1)^{l-i}\cdot h_{M-s-i},$ for arbitrary but fixed integers $l$ and $s$. We will further discuss the filter operators in the context of NDWT later in this section.

\begin{figure*}
\begin{center}
        \subfigure[]{
            \includegraphics[width=0.45 \columnwidth]{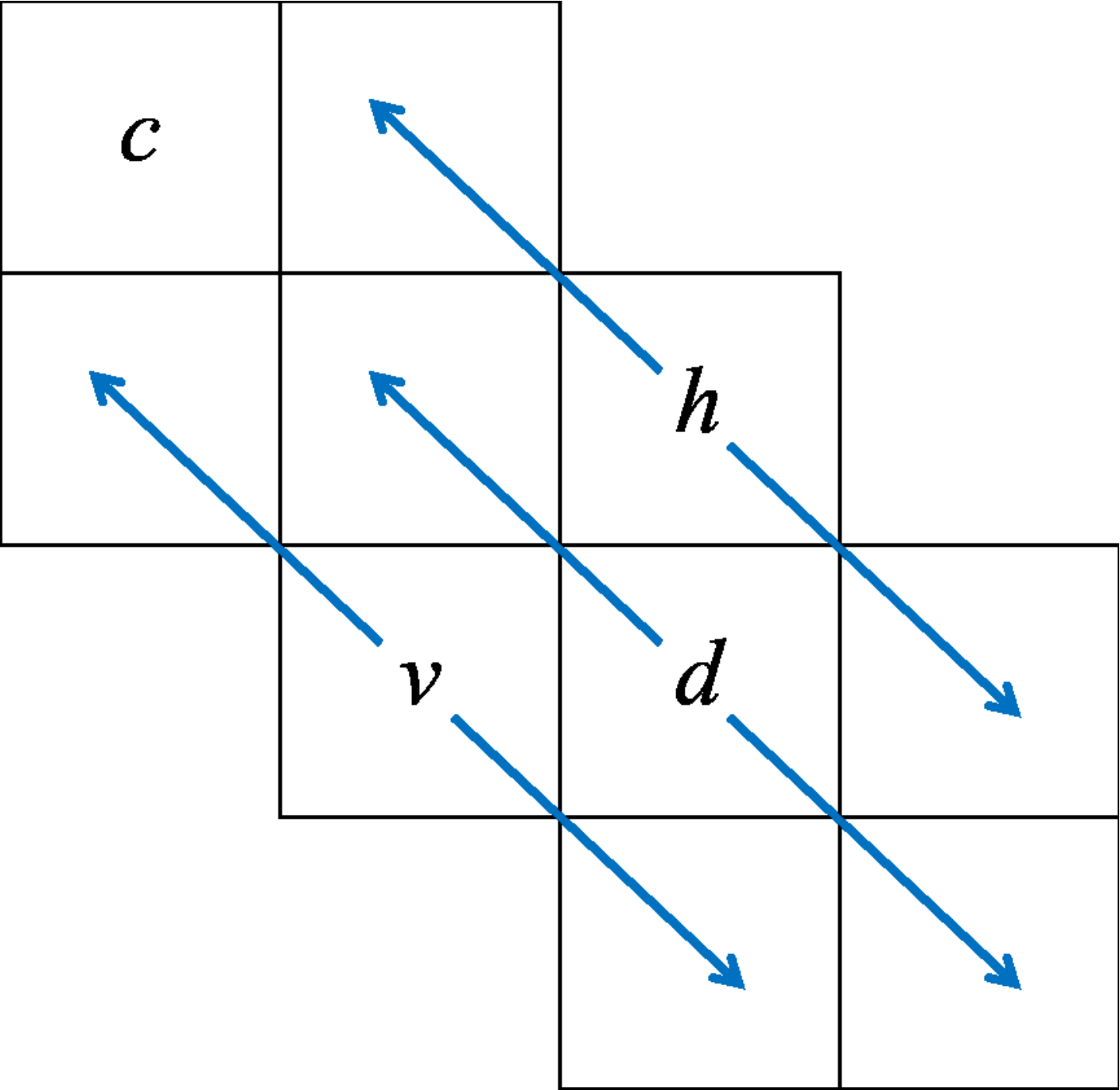}
        }
        \subfigure[]{
            \includegraphics[width=0.45 \columnwidth]{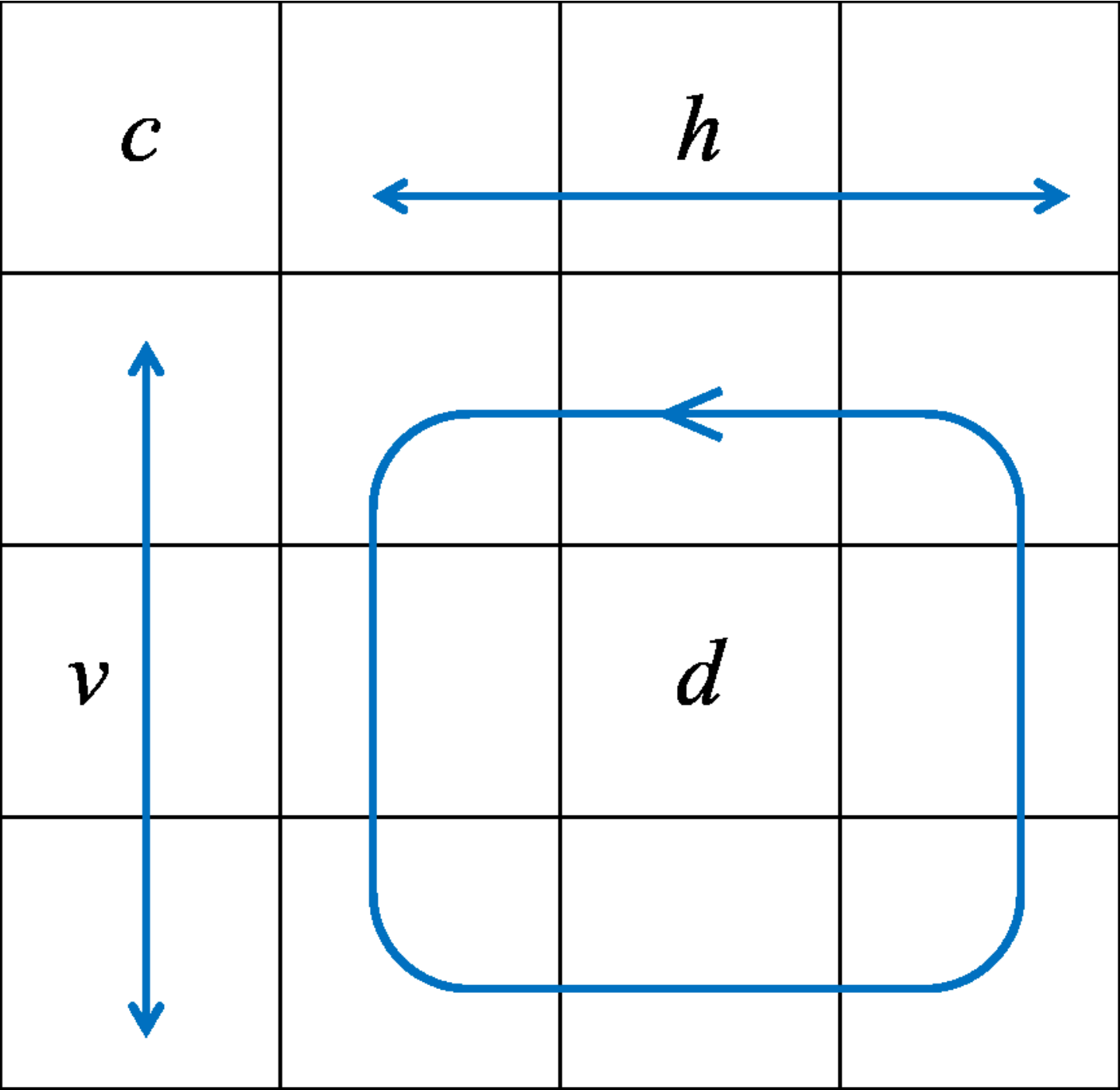}
        }
\end{center}
\caption{Locations of four types of wavelet coefficients in the tessellation of 3-level decomposition with the standard and scale-mixing 2-D NDWTs. Different types of coefficients are defined in (\ref{eq:coefsnd2d}).}
\label{fig:wavCoef}
\end{figure*}

Expanding on the 1-D definitions, we overview a 2-D NDWT of $f(x,y)$, where $(x,y)\in \mathbb{R}^2$. Several versions of 2-D NDWT exist but we focus on the standard and a scale-mixing versions.  For the standard 2-D NDWT, the wavelet atoms are
\ba
\phi_{J_0; k_1,k_2} (x,y)& = & 2^{ J_0} \phi(2^{J_0} (x - k_1)) \phi(2^{J_0} (y - k_2)),\\
\psi_{j; k_1,k_2}^{(h)}(x,y)& = & 2^{j} \phi(2^{ j} (x - k_1))\psi(2^{ j} (y - k_2)),\\
\psi_{j; k_1,k_2}^{(v)}(x,y)& = & 2^{j} \psi(2^{ j} (x - k_1))\phi(2^{ j} (y - k_2)),\\
\psi_{j; k_1,k_2}^{(d)}(x,y)& = & 2^{j} \psi(2^{ j} (x - k_1))\psi(2^{ j} (y - k_2)),
\ea
where $(k_1,k_2)$ is the location pair, and $j = J_0, \dots, J-1$ is the scale. The depth of the transform is $p=J-1-J_0.$
The wavelet coefficients of $f(x,y)$ are calculated as
\ba
c_{J_0;k_1,k_2} & = & 2^{J_0}  \iint f(x,y)
 \phi_{J_0;k_1,k_2} (x,y) \; dxdy, \\
d_{ j;k_1,k_2}^{(i)}& = & 2^{ j}  \iint f(x,y)
 \psi_{j;k_1,k_2}^{(i)} ( x,y) \; dxdy,
\ea
where $J_0$ is the coarsest decomposition level, and $i \in \{h, v, d\}$ indicates the ``orientation'' of detail
coefficients as horizontal, vertical, and diagonal (e.g., \cite{vidakovic2009}, p. 155).
The tessellation to a standard 2-D NDWT is presented in Figure \ref{fig:wavCoef}(a).

For the scale-mixing 2-D NDWT, the wavelet atoms are
\ba
\phi_{J_{01},J_{02}; \bm k} (x, y) & = & 2^{(J_{01}+J_{02})/2} \phi(2^{J_{01}} (x - k_1)) \phi(2^{J_{02}} (y - k_2)),\\
\psi_{J_{01}, j_2; \bm k}(x, y) & = & 2^{(J_{01}+j_2)/2} \phi(2^{J_{01}} (x - k_1))\psi(2^{ j_2} (y - k_2)),\\
\psi_{j_1, J_{02}; \bm k}(x, y)  & = & 2^{(j_1+J_{02})/2} \psi(2^{ j_1} (x - k_1))\phi(2^{ J_{02}} (y - k_2)),\\
\psi_{j_1, j_2; \bm k}(x, y) & = & 2^{(j_1+j_2)/2} \psi(2^{ j_1} (x - k_1))\psi(2^{ j_2} (y - k_2)),
\ea
where $J_{01}$  and $J_{02}$ are coarsest levels, $j_1 \geq J_{01}; j_2 \geq J_{02},$  and $\bm k = (k_1, k_2).$
 As a result, we obtain wavelet coefficients for $f(x,y)$ from the scale-mixing NDWT as
\be
\label{eq:coefsnd2d}
c_{ J_{01},J_{02}; \bm k } & = &  \iint   f(x,y)
\phi_{J_{01},J_{02}; \bm k} (x, y)\; dxdy, \nonumber \\
h_{ J_{01},j_{2} ; \bm k } & = &  \iint   f(x,y)
\psi_{J_{01}, j_2; \bm k}(x, y)\; dxdy,\\
v_{ j_{1},J_{02}; \bm k } & = &  \iint  f(x,y)
\psi_{j_1, J_{02}; \bm k}(x, y)  \; dxdy, \nonumber \\
d_{ j_1,j_2;\bm k } & = &  \iint   f(x,y)
\psi_{j_1, j_2; \bm k}(x, y) \; dxdy. \nonumber
\ee
 Notice that in the standard NDWT, we use common $j$ to denote a scale, while in the scale-mixing NDWT, we use
a pair $(j_1, j_2),$ which indicates that two scales are mixed.  Figure \ref{fig:wavCoef}(b) illustrates the
 tessellation of coefficients of scale-mixing 2-D NDWT. In Section \ref{sec:2dsmndwt} we will refer
 to coefficients from (\ref{eq:coefsnd2d}) as $c$-,$h$-, $v$-, and $d$-type coefficients.

%

While the functional series involving wavelet and scaling functions as decomposing atoms is an established mathematical framework for describing the NDWT,
we provide an alternative description of NDWT using convolution operators (\cite{nason1995stationary}, \cite{strang1996wavelets}, \cite{vidakovic2009}  ).
Such a description is preferred for discrete inputs.

Let $[\uparrow 2]$ denote the upsampling of a given sequence by inserting a zero between every
two neighboring elements of a sequence. We define the dilations of wavelet filters $h$ and $g$ as
\begin{align}
\label{eq:filters}
h^{[0]} = &  h, & g^{[0]}&= g\\ \nonumber
h^{[r]} = & [\uparrow 2] h^{[r-1]}, & g^{[r]}&=[\uparrow 2] g^{[r-1]}.
\end{align}
Inserting zeros between each element of filters $h^{[r-1]}$ and $g^{[r-1]}$ creates holes
({\it trous}, in French), which is why this approach is sometimes called Algorithm {\'a} Trous,
see \cite{shensa1992}.

A non-decimated wavelet transform is completed by applying convolution operators, ${\bf H}^{[j]}$ and ${\bf G}^{[j]}$, which come from dilated filters $h^{[r]}$ and $g^{[r]}$ in sequence.
\begin{figure}[h]
\centering
\includegraphics [scale=0.5 , clip]{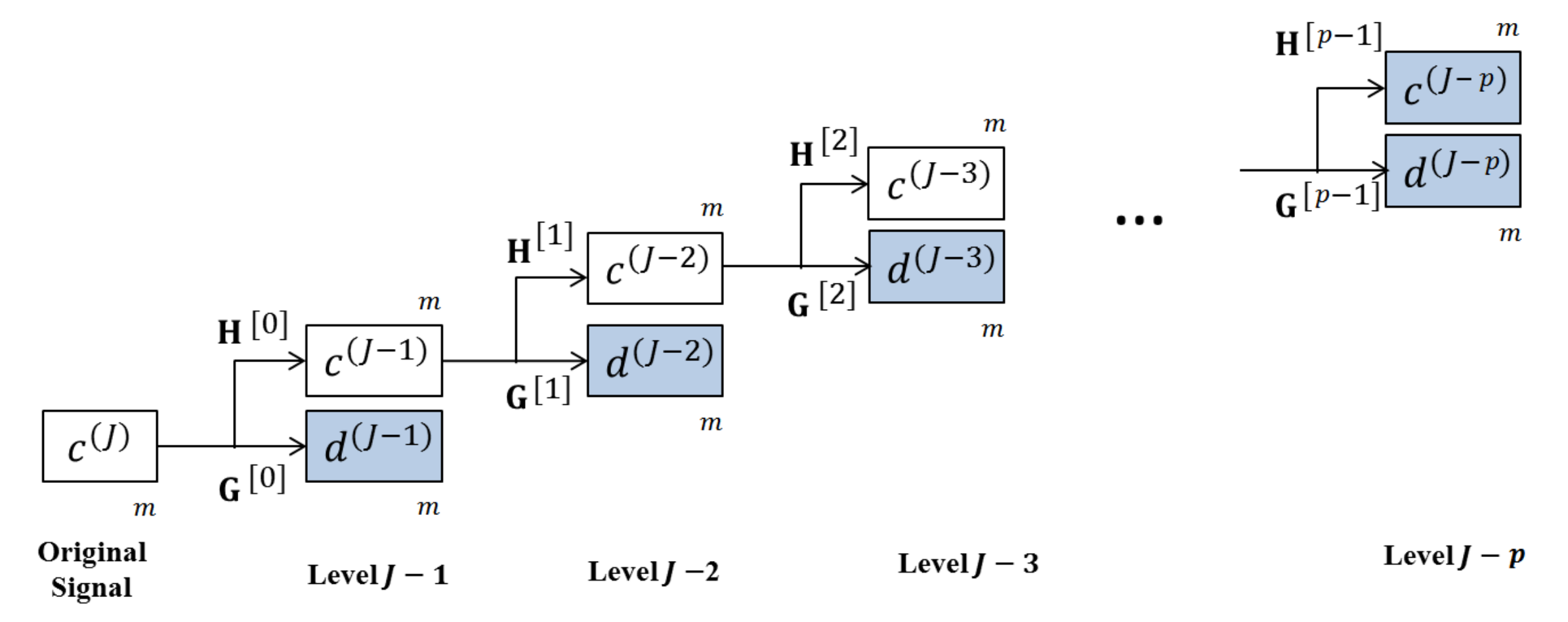}
\caption{ Graphical explanation of the NDWT process. Given signal $a^{J}$ of size $m$, we decompose the signal to $p+1$ multi-resolution subspaces that include  $p$ levels of detail coefficients and one level of scaling coefficients, resulting in a set of coefficient vectors, ${\bm d}^{(J-1)}, \bm{d}^{(J-2)}, \dots,  \bm{d}^{(J-p)},$ and  $\bm c^{(J-p)}$ in shaded blocks.} \label{fig_NDWTG}
\end{figure}
Detail and coarse coefficients generated from each level have an identical size, $m$, which is the same as that of the original signal. To obtain coefficients at decomposition level $J-j$, where $j \in \{1, 2, \dots, p\} $, we repeatedly apply convolution operators to a coarse coefficient vector from the previous decomposition level, $J-j+1$
\ba
\bm c^{(J-j)} & = & {\bf H}^{[j-1]} \bm c^{(J-j+1)} \\
\bm d^{(J-j)} & = & {\bf G}^{[j-1]} \bm c^{(J-j+1)},
\ea
where ${\bf H}^{[j-1]}$ and ${\bf G}^{[j-1]}$ are filter operators that perform low- and high-pass filtering using quadrature mirror filters $h^{[j-1]}$ and $g^{[j-1]}$, respectively. The NDWT is the result of repeated applications of two filter operators, ${\bf H}^{[j]}$ and ${\bf G}^{[j]}$.  Operators (${\bf H}^{[j]}$, ${\bf G}^{[j]}$) do not have an orthogonality property, so to obtain such a property, we utilize two additional operators $\mathcal{D}_0$ and $\mathcal{D}_1$, which perform decimation by selecting every even and odd member of an input signal.  An example of the use of the decimation operator $\mathcal{D}$ with a signal $x$ is
\ba
(\mathcal{D}_0 x)_i &=& x_{2i}, \\
(\mathcal{D}_1 x)_i&=& x_{2i+1},
\ea
where $i$ indicates the position of an element in the signal $x$.  We apply ($\mathcal{D}_0{\bf H}^{[j-1]}$, $\mathcal{D }_0{\bf G}^{[j-1]}$) and ($\mathcal{D}_1{\bf H}^{[j-1]}$, $\mathcal{D}_1{\bf G}^{[j-1]}$) to a given signal and obtain the even and odd elements of NDWT wavelet coefficient vectors, $\bm c^{(J-j)}$ and $\bm d^{(J-j)}$, respectively. Thus, equation (\ref{eq:filters}) is, in fact, performed as the following process
\ba
(\bm c^{(J-j)})_{2i} & = & \mathcal{D}_0{\bf H}^{[j-1]} \bm c^{(J-j+1)} \\
(\bm c^{(J-j)})_{2i+1} & = & \mathcal{D}_1{\bf H}^{[j-1]} \bm c^{(J-j+1)} \\
(\bm d^{(J-j)})_{2i} & = & \mathcal{D}_0{\bf G}^{[j-1]} \bm c^{(J-j+1)} \\
(\bm d^{(J-j)})_{2i+1} & = & \mathcal{D}_1{\bf G}^{[j-1]} \bm c^{(J-j+1)}.
\ea
We apply the filtering twice at the even and odd positions for each decomposition level, so a shift does not affect transformation results, which means that the NDWT is time-invariant. Such time-invariance property of the NDWT yields a smaller mean  squared error and reduces the Gibbs phenomenon in de-noising applications \citep{coifman1995translation}. However, the violation of variance preservation in the NDWT complicates the signal reconstruction. In the following section we will discuss how to perform lossless reconstruction of an original image using a matrix-based NDWT.\\


\section{Matrix Formulation of NDWT}
\label{sec:mf2d}
In this section, we translate multiple convolutions in the NDWT into a simple matrix multiplication. In Mallat's algorithm, scaling and wavelet functions are convolved in a cascade. Instead of performing convolutions with wavelet and scaling functions, we formulate the NDWT as matrix multiplication.  We simplify the cascade algorithm as follows. With filtering matrices, Mallat's cascade algorithm is implicit in repeated matrix multiplications of low- and high-pass filter matrices, ($H$) and ($G$), respectively. The following matrices
illustrate combination  of component filter matrices, to achieve transforms of depth 1, 2, and 3.

\begin{align*}
W_{m}^{(1)} = \begin{bmatrix}
        H_1\\[0.3em]
       G_1
     \end{bmatrix}_{[2m\times m]} ,
     W_{m}^{(2)} = \begin{bmatrix}
     \begin{bmatrix}
           H_2\\[0.3em]
           G_2
     \end{bmatrix} \cdot  H_1\\[0.3em]
           G_1
     \end{bmatrix}_{[3m\times m]}  ,
     W_{m}^{(3)} = \begin{bmatrix}
     \begin{bmatrix}
          \begin{bmatrix}
           H_3\\[0.3em]
           G_3
          \end{bmatrix} \cdot  H_2\\[0.3em]
           G_2
          \end{bmatrix} \cdot H_1\\[0.3em]
           G_1
          \end{bmatrix}_{[4m\times m]} , \dots
\end{align*}\\
Filter matrices $\begin{bmatrix} H_p &  G_p  \end{bmatrix}^T $ as submatrices of $W_{m}^{(p)}$ are formed by simple rules. The sizes of $H_p$ and $G_p$ for $p\in \{1,2,\dots \}$ are the same, $m \times m,$ and their entries at the position $(i,j)$ are
\ba
a_{ij}= & \frac{1}{\sqrt{2}} h^{[p-1]}_s, \; & s=N + i -j\; modulo \; m  \\
b_{ij}= & \frac{1}{\sqrt{2}} (-1)^s h^{[p-1]}_{N+1-s}, \; & s=N + i -j\; modulo \; m,\\
\ea
respectively, where $N$ is a shift parameter and $h^{[p-1]}_s$ is the $s^{th}$ element of a dilated wavelet filter $h$ with $p-1$ zeros in between the original components $(h_1, h_2, \dots, h_u)$,
\ba
h^{[p-1]}=(h_1, \overbrace{0, \dots, 0}^\text{$p-1$}, h_2, \overbrace{0, \dots, 0}^\text{$p-1$}, h_3,\dots, \overbrace{0, \dots, 0}^\text{$p-1$}, h_u)\\
\ea
For example, $h^{[p-1]}_1=h_1$, $h^{[p-1]}_{p+1}=h_2$, \dots, and, $h^{[p-1]}_{p(u-1)+1}=h_u$. Following such construction rules, ${W}_{m}^{(p)}$ becomes a matrix of size $\big(m (p+1) \, \times \, m\big)$ consisting of $p+1$ stacked submatrices of size $[m \times m]$.  The NDWT matrix formed in the described process is not normalized
and signal reconstruction cannot be done by using its transpose only.
Indeed, in terms of Mallat's algorithm, for the inverse transform, at each step the multiplication by 1/2 is needed
for perfect reconstruction  \cite[see][Proposition 5.6]{mallatbook}.

Thus, we construct a diagonal weight matrix that rescales the square submatrices comprising
the NDWT matrix, to be used when performing the inverse transform.
 The weight matrix for ${W}_{m}^{(p)}$ has size $(m(p+1)\times m(p+1))$ and is defined as
\ba
 {T}_{m}^{(p)}=\mbox{diag}(\overbrace{1/2^p, \dots,1/2^{p}}^\text{$2m$},\overbrace{1/2^{p-1},\dots,1/2^{p-1}}^\text{$ m$},\overbrace{1/2^{p-2},\dots,1/2^{p-2}}^\text{ $m$},\dots, \overbrace{1/2, \dots,1/2}^\text{$m$}). \\
\ea

A 1-D signal $\bm y$ of size $[m\times 1]$ is transformed in a $p$-level decomposition to a vector $\bm d$
by multiplication by wavelet matrix ${W}_{m}^{(p)}.$
The original signal is then reconstructed by multiplying $\bm d$ by  $({W}_{m}^{(p)})'$ rescaled by the weight matrix ${T}_{m}^{(p)}$. \\

\be
\label{eq:1d}
\bm d&=& {W}_{m}^{(p)}\times \bm y_{[m\times 1]} \nonumber  \\
\bm y&=&  ({W}_{m}^{(p)})'\times {T}_{m}^{(p)}\times \bm d,
\ee
where $p$ and $m$ are arbitrary.

 Note that $  ({W}_{m}^{(p)})'  \times  {W}_{m}^{(p)}   \ne I_m. $ On the other hand,
 column vectors of matrix $V_m^{(p)} = \left(T_m^{(p)}\right)^{1/2} {W}_{m}^{(p)}$ form an orthonormal set, that is,
 \be
 \label{eq:VnotW}
 (V_m^{(p)})' \times  V_m^{(p)} = I_m
 \ee
 The product $V_m^{(p)}  \times (V_m^{(p)})' $ cannot be an identity matrix, but
 \ba
 \sum_i \left( V_m^{(p)}  \times (V_m^{(p)})' \right)_{ij} =
 \left(\sum_j \left( V_m^{(p)}  \times (V_m^{(p)})' \right)_{ij}\right)'=
 \left[{\bm 1}_m, {\bm 0}_{p m} \right],
 \ea
 where $\left[{\bm 1}_m, {\bm 0}_{p m} \right]$ is a row vector consisting of $m$ ones followed by the $p m$ zeros.

 Since $I_m = (V_m^{(p)})' \times  V_m^{(p)}  = ({W}_{m}^{(p)})' \times T_m^{(p)} \times  {W}_{m}^{(p)} $,
 the perfect reconstruction is achieved by $({W}_{m}^{(p)})'\times {T}_{m}^{(p)}$ applied on the vector
 transformed by ${W}_{m}^{(p)}$, as in (\ref{eq:1d}).

 Although transformation by $V_m^{(p)}$ looks more natural because of (\ref{eq:VnotW}),
 the scaling of wavelet coefficients  when transformed by $ V_m^{(p)}$  is not matching the correct scaling
 produced by Mallat's algorithm, or equivalently, by integrals in (\ref{eq:ndseries}).  The correct scaling of wavelet coefficients is important
 in applications involving regularity assessment of signals and images, as we will see
 in the mammogram example from Section \ref{sec:twoexamples}.


\subsection{Scale-Mixing 2-D NDWT}
\label{sec:2dsmndwt}
A 2-D signal $\bm A$ of size $[m \times n]$ for $p_1$- and $p_2$-level decomposition along rows and columns, respectively,
 is obtained by NDWT matrix multiplication from the left and its transpose from the right.
 The   transform results in a 2-D signal $\bm B$ of size  $(p_1 + 1) m \times (p_2 + 1) n$.
The inverse transform applies the rescaling matrices ${T}_{m}^{(p_1)}$ and ${T}_{n}^{(p_2)}$
on the corresponding NDWT matrices,
\be
\label{eq:2d}
\bm B&=&   {W}_{m}^{(p_1)}  \times \bm A_{[m\times n]}\times ({W}_{n}^{(p_2)})'   \nonumber \\
\bm A&=&  ({W}_{m}^{(p_1)})'\times  {T}_{m}^{(p_1)}\times \bm B \times  {T}_{n}^{(p_2)}\times  {W}_{n}^{(p_2)},
\ee
Here   $p_1$, $p_2$, $m$, and $n$ can take any integer value, and ${W}_{m}^{(p_1)}$ and ${W}_{m}^{(p_2)}$ could be constructed using possibly different wavelet filters.

\begin{figure}[ht!]
\centering
\includegraphics [scale=0.35, clip]{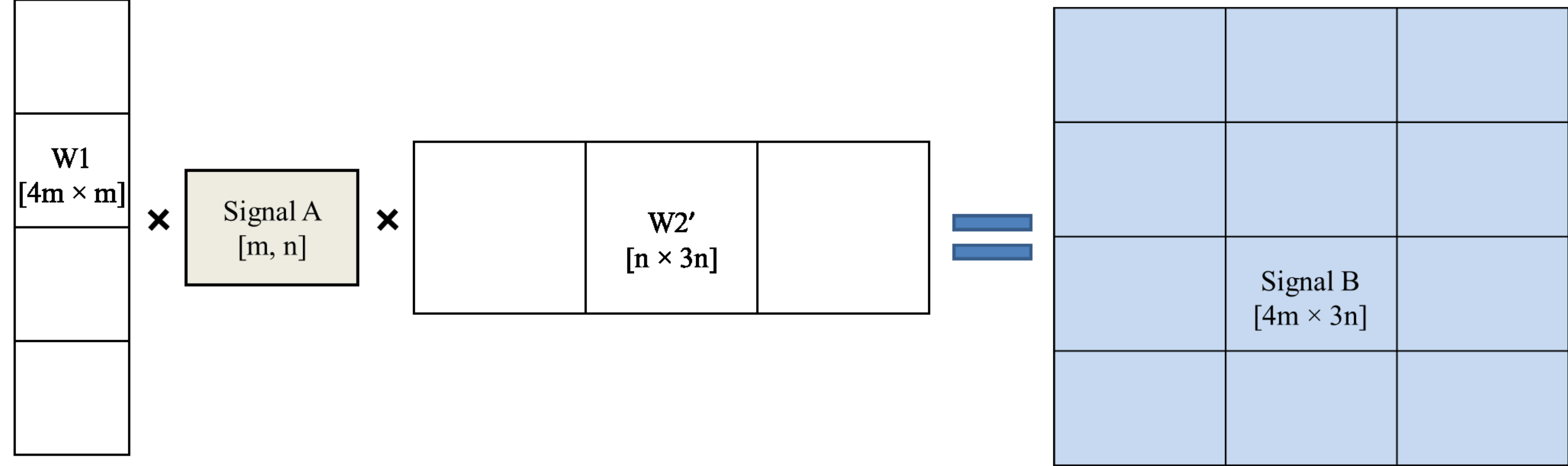}
\caption{Graphical illustration of a 2-D NDWT scale-mixing transform with $3$-levels along  the columns and $2$-levels
 along the rows.    The NDWT matrices $W_1$ and $W_2$ can be constructed by possibly different wavelet filters.} \label{fig_MatGrid}
\end{figure}

\begin{figure*}
\begin{center}
        \subfigure[]{
            \includegraphics[width=0.45 \columnwidth]{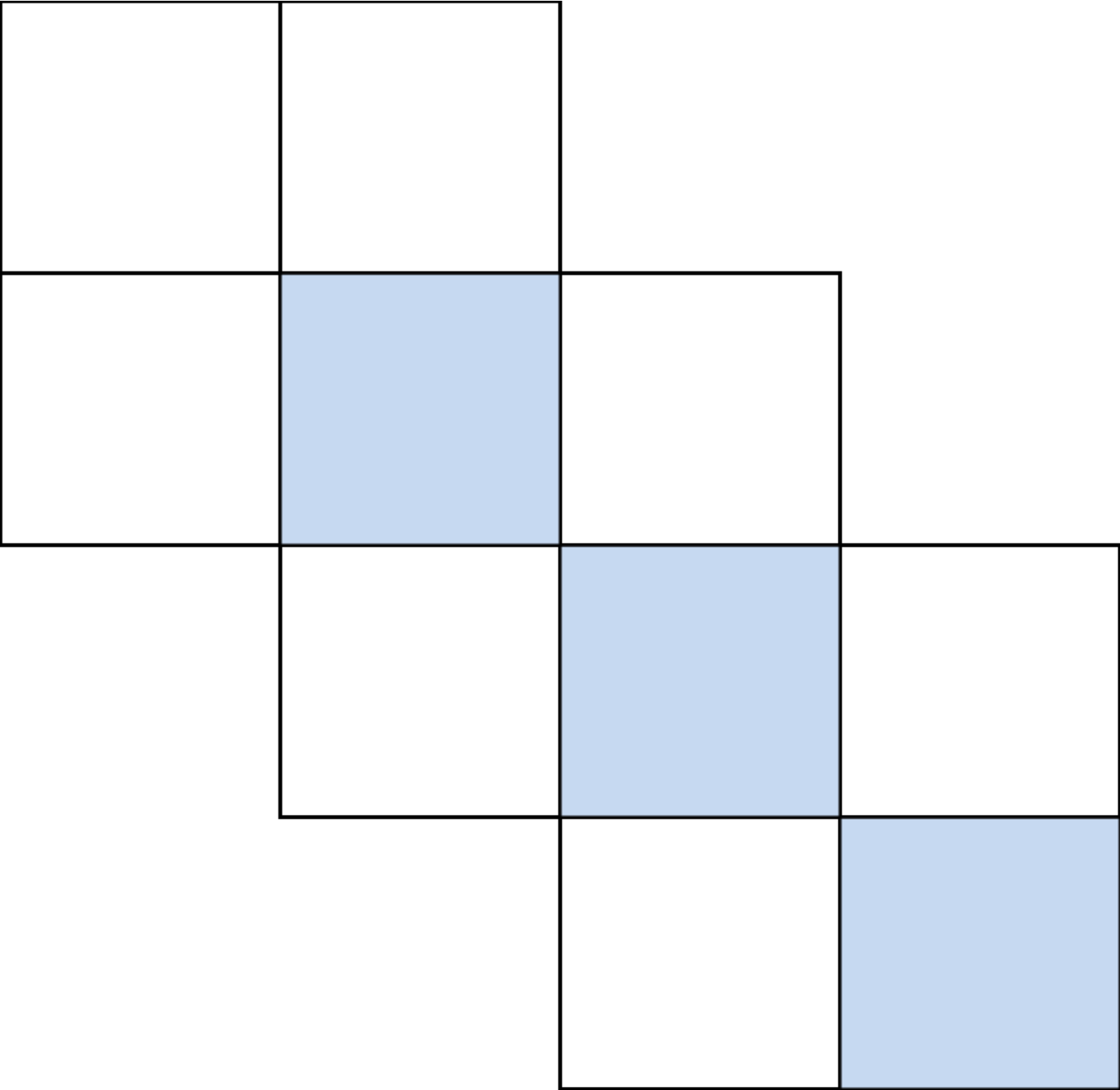}
        }
        \subfigure[]{
            \includegraphics[width=0.45 \columnwidth]{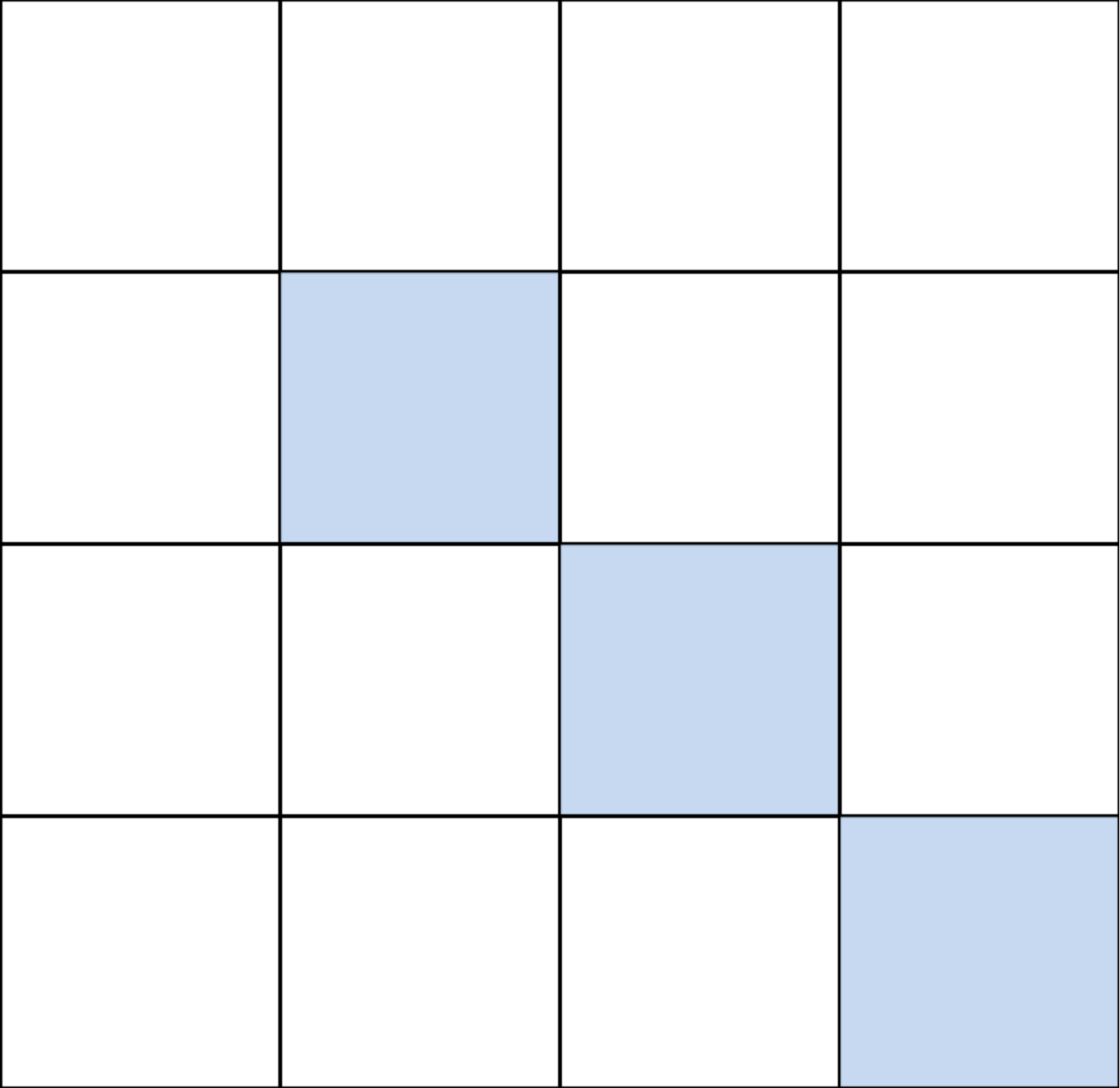}
        }
\end{center}
\caption{Tesselation of 3-level decomposition with standard (left) and scale-mixing (right) 2-D NDWT. Shadded areas correspond to $d$-type wavelet coefficients}
\label{fig:compression}
\end{figure*}

One of the advantages of the scale-mixing 2-D NDWT is its superior compressibility.
 Wavelet transforms act as approximate Karhunen-Lo\`eve transforms and
   compressibility in the wavelet domain is beneficial in tasks wavelet-based data compression and  denoising.
When an image possesses a certain degree of smoothness, the coefficients corresponding to
diagonal decomposition atoms [$d$-coefficients in (\ref{eq:coefsnd2d})] tend to be
smaller in magnitude compared to the $c$-, $v$- or $h$-type coefficients in (\ref{eq:coefsnd2d}).
 As an example, consider  performing a $p$-level decomposition of a 2-D image of size $[m \times n]$ with the both NDWT matrix (as scale-mixing) and standard 2-D NDWT. The compressibility of transform can be defined as the proportion of diagonal-type coefficients divided by the total number of wavelet coefficients. As we mentioned before, $d$-coefficients correspond to decomposing atoms consisting of two wavelet functions, while the atoms of $c$-, $v$- or $h$-type coefficients contain at least one scaling function.
 In the scale-mixing NDWT of depth $p$,   $p^2 m n  / ( (p+1)^2 m n)$ is the proportion of $d$-type coefficients,
   while in the standard   2-D NDWT this proportion is $p  m n/  ((3p+1) m n)$ (see Figure \ref{fig:compression}). The former is always greater than the later, except when $p=1$, in which case the two proportions coincide. Thus, the scale-mixing 2-D NDWT tends to be more compressive compared to the standard 2-D NDWT.

\begin{figure*}
\begin{center}
        \subfigure[]{
            \includegraphics[width=0.5 \columnwidth, height = 0.25 \columnwidth]{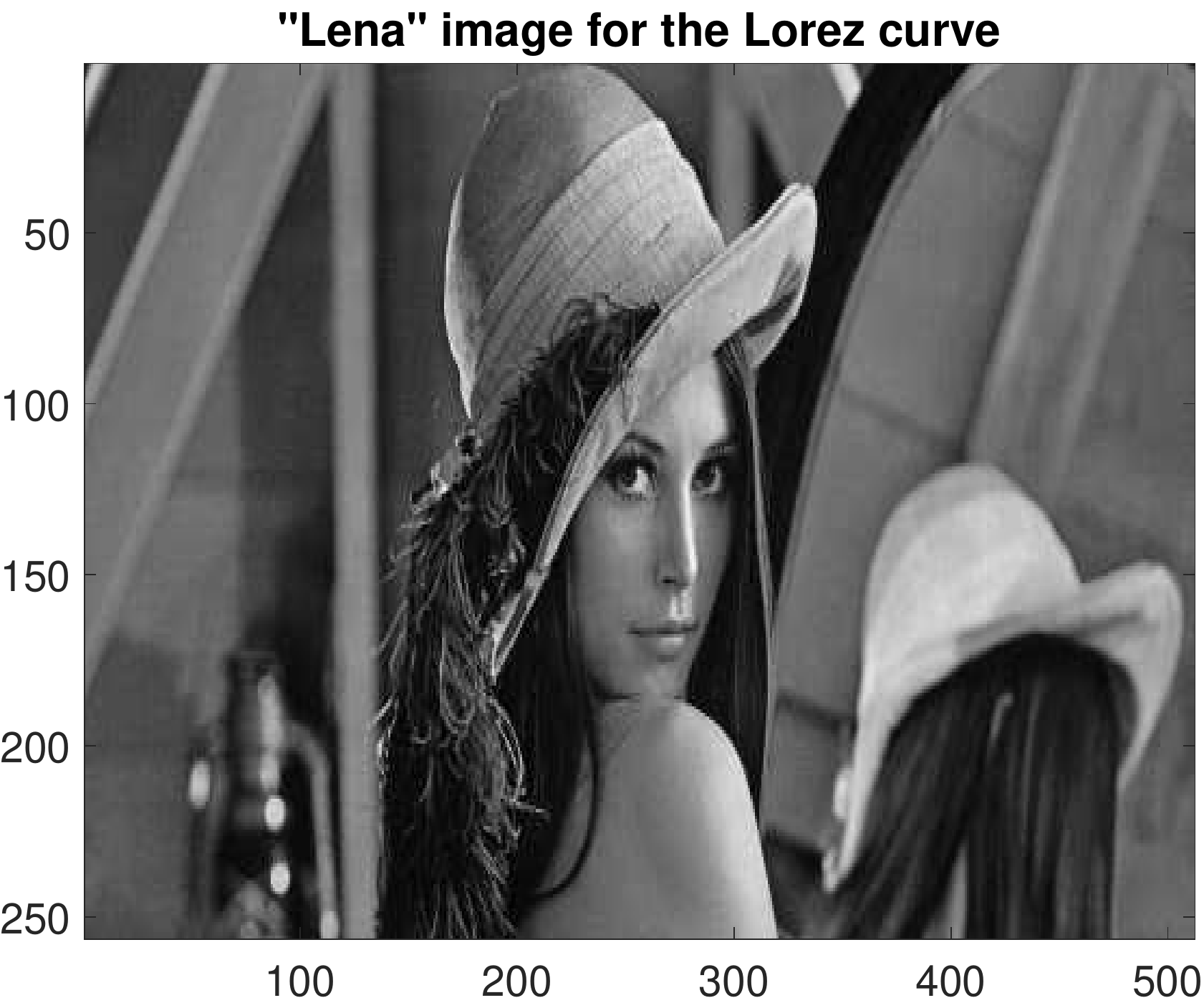}
        }
        \subfigure[]{
            \includegraphics[width=0.4 \columnwidth, height = 0.25 \columnwidth]{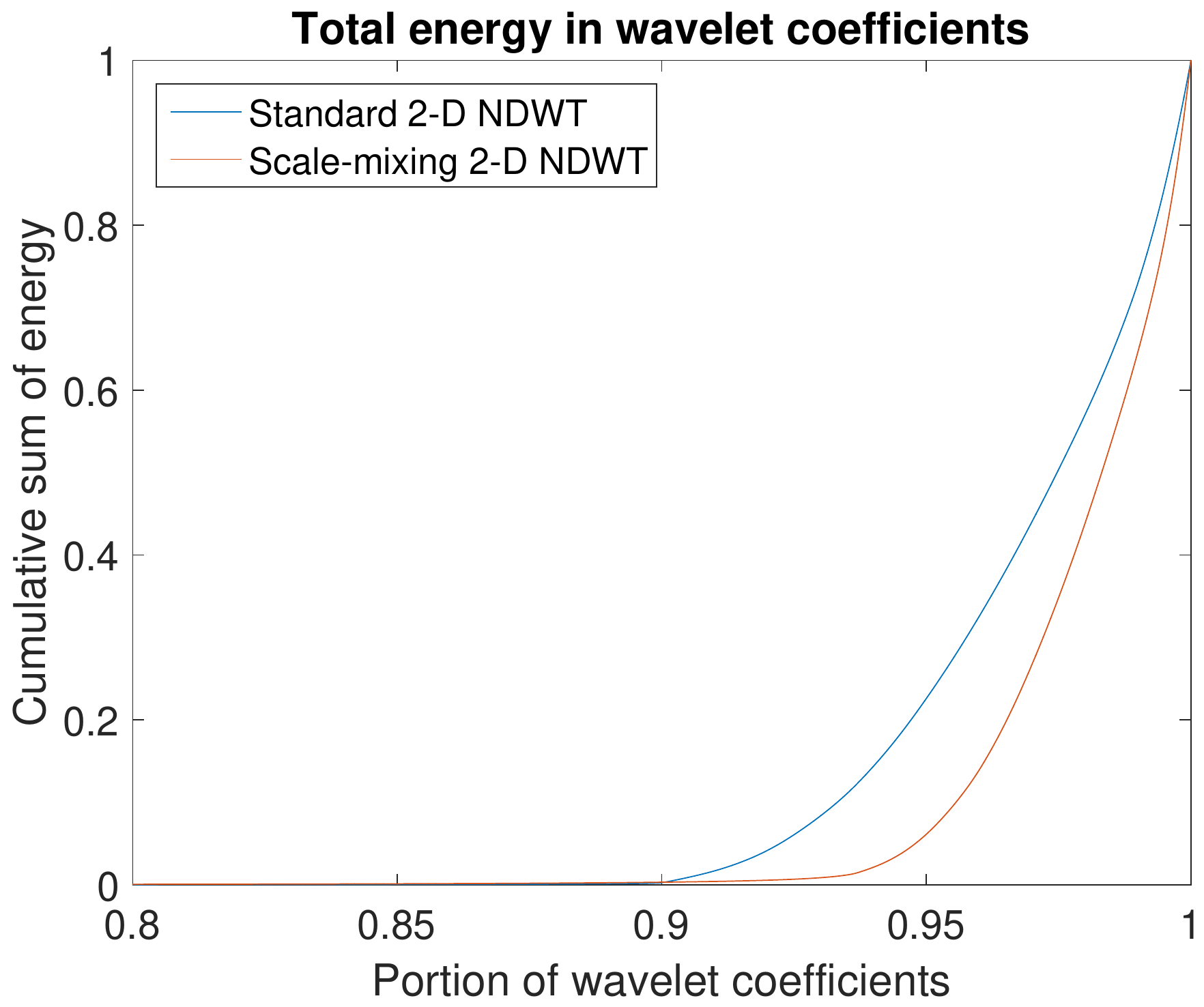}
        }
\end{center}
\caption{Image in panel (a) is transformed with both scale-mixing and standard 2-D NDWT into 3 decomposition levels based on a Haar wavelet filter. A detail of the Lorenz curve in panel (b) indicates that the 2-D scale-mixing NDWT compresses the information in image more efficiently compared to the traditional transform.}
\label{fig:lorenz}
\end{figure*}

As an illustration, we transform a noiseless ``Lena'' image of size $256 \times 512$ (Figure \ref{fig:lorenz}(a)) with  both the standard and scale-mixing 2-D NDWT in a 3-level decomposition using the Haar wavelet. To compare the compressibility, we calculate and contrast  Lorenz curves.
For the Lorenz curve, we normalize all squared wavelet coefficients as $p_k=d_{k}^2/(\sum_i d_{i}^2)$, sort $p_k$ in an increasing order, and obtain the cumulative sum of sorted $p_k$. This cumulative sum of (normalized) energy
for the two transforms is shown Figure \ref{fig:lorenz}b ). The curves are plotted against the portion of wavelet coefficients
 used in the cumulative sum.
 At top right corner of Figure \ref{fig:lorenz}b, the curves meet, since for both curves $\sum p_i =1$. However, the blue curve (standard NDWT) uniformly dominates the red curve (scale-mixing NDWT). This means that the compressibility of the scale-mixing NDWT is higher. In simple terms, the scale-mixing NDWT requires smaller portion of the wavelet coefficients to preserve the same relative ``energy.'' To numerically quantify this compressibility, we think of $p_k$'s as the probabilities and calculate entropies of their distributions. Calculating the normalized Shannon entropy, $(\sum_{i=1}^n p_i \log p_i)/\log n $, we obtain 0.7994 for the scale-mixing NDWT and 0.8196 for the standard NDWT. The scale-mixing NDWT has lower entropy, which confirms its superior compressibility. Although demonstrated here only on ``Lena" image, this superiority in compression
  for scale-mixing transforms holds generally,  see \cite{remenyi2014image}.

 \section{Computational Efficiency of the NDWT Matrix}
Next we discuss several features of NDWT matrix, so that users are aware of its advantages as well as limitations.

Principal advantages of a NDWT matrix are compressibility,  computational speed, and flexiblity in size of an input signal.
  We already discussed the better compressibility when NDWT matrices are used for 2-D scale-mixing  transforms.

 Next, we compare  the computation time of the matrix-based NDWT to that of the convolution-based NDWT.
The NDWT matrix performs a transform faster than the convolution-based NDWT.
This statement is conditional on the software used for the computation. We used {MATLAB} version 8.6.0.267246 (R2015b, 64-bit) on a laptop with quad-core CPU running at 1,200 MHz with 8GB of RAM.

At first glance, improving the speed of calculation by using matrix multiplication over convolutions
looks counterintuitive. The asymptotic computational complexity for convolutions is much lower than the complexity of matrix multiplication. The NDWT based on Mallat's algorithm has calculational complexity
of $O( n \log n )$, while the (na\"ive) matrix
 multiplication has the complexity of $O(n^3).$ The complexity of matrix multiplication could be improved by
 the \cite{legall2014} algorithm to $O(n^{2.3729}),$ with a theoretical lower bound of $O(n^2 \log n),$
 still inferior to convolutions.
 However, the ``devil is in the constants.'' For signals of moderate size, the calculational overhead that manages repeated filtering operations in convolution-based approach slows down the computation and direct matrix multiplication
 turns out to be faster.

\begin{figure}[h]
\centering
\includegraphics [scale=0.45 , clip]{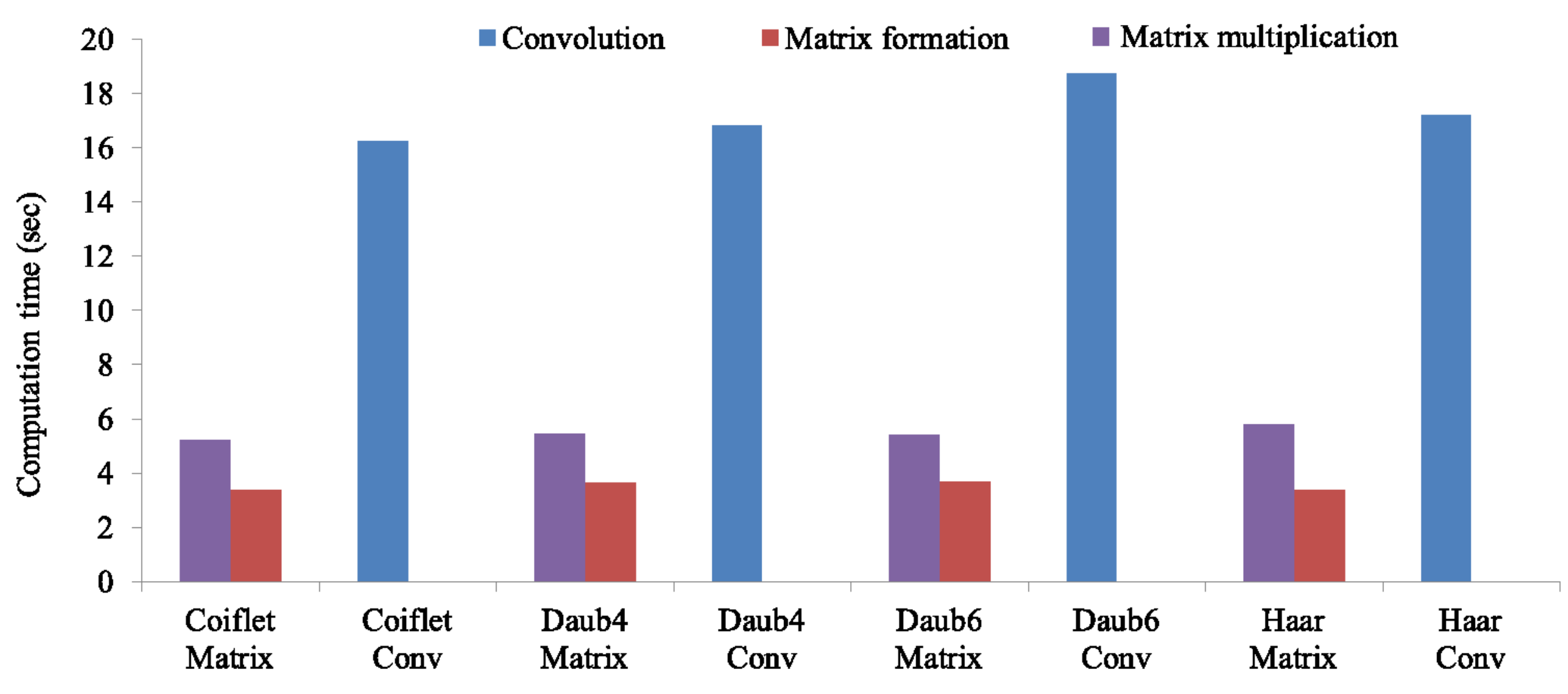}
\caption{Average computation time (in seconds) to perform the matrix-based and convolution-based NDWT for 8-level decompositions along both rows and columns using Coiflet, Daubechies 4, Daubechies 6, and Haar wavelets. The size of inputs is $2^{10} \times 2^{10}$ and the computation time is averaged over $100$ repetitions.} \label{compt}
\end{figure}
As an illustration, we simulated 100 2-D fractional Brownian fields (fBf) of size ($2^{10} \times 2^{10}$) with the Hurst exponent $H=0.5$ and performed the eight-level decomposition NDWT with four wavelets:  Haar, Daubechies (4 and 6 tabs), and Coiflet.  For the NDWT of a single signal, the computation time of the matrix-based NDWT was on average of 9.02 seconds while that of the convolution-based NDWT was on average of 17.26 seconds.  In addition, about 40 \% of the computation time of the matrix-based NDWT was spent on constructing an NDWT matrix that could be used repeatedly in simulation for the same type of NDWT once generated. Thus, as the NDWT is repeated on the input signals of the same size using the same wavelet filter, the difference in computation time becomes even greater. For a 1-D signal transform, the matrix-based NDWT is approximately twice as fast as the convolution-based NDWT under the given conditions, but this factor increases to three for the NDWT of 100 signals having the same size and transformed using the same matrix.

While NDWT matrices reduce the computation time by storing all entries of the matrices used in convolution for each decomposition level in a single matrix, such property can limit the usage of NDWT matrices. When the size of an input is large, a computer with standard specifications may not have enough memory to store a NDWT matrix of appropriate size. 
This issue affects mostly the cases of 1-D signals. For 2-D transforms, if the computer can store an image,
it can most likely store the NDWT matrix, since the matrix is only $(p + 1)$ times larger,
and $p$ is typically small. 
To find a limit on the size of an 1-D input, we repeatedly constructed NDWT matrices for one-level decomposition increasing the size of an input by $500$ in each trial. We found that as the size of an input signal exceeded $35,000$, matrix construction was not possible because of  limited memory capacity.
\begin{figure}[h]
\centering
\includegraphics [scale=0.45 , clip]{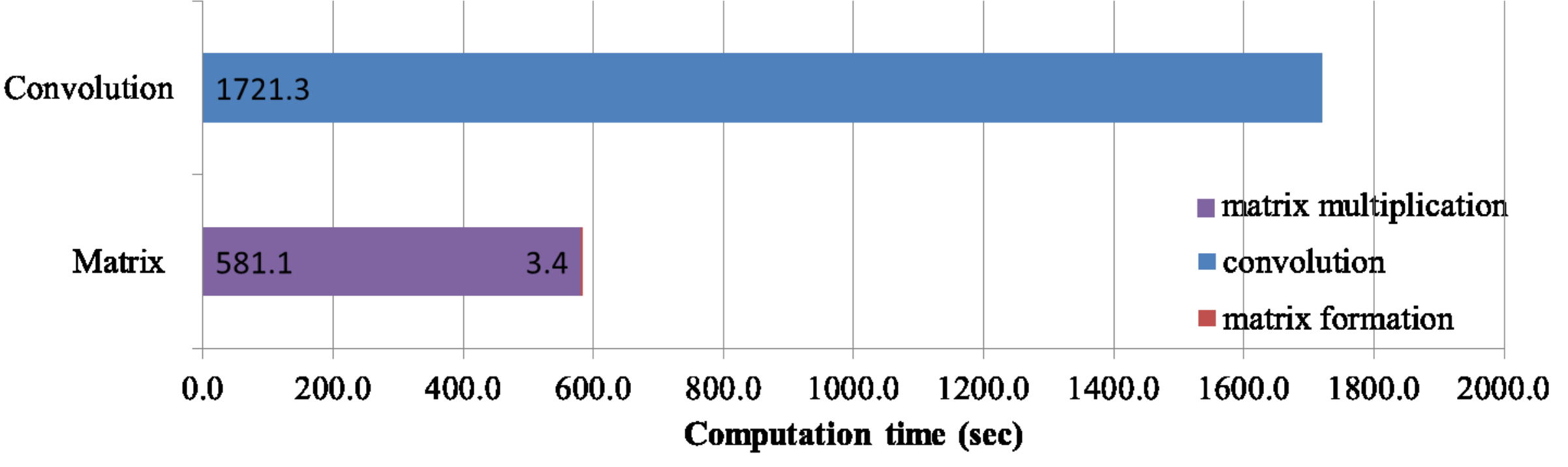}
\caption{Computation time (in seconds) of the matrix- and convolution-based NDWT for 8-level decomposition evaluated for 100 2-D signals of the size ($1,024 \times 1,024$) using Haar wavelet. The matrix was pre-constructed to perform the same type of transform.}
\label{graph:compt}
\end{figure}

The matrix-based   NDWT can be applied to signals of a non-dyadic length and for 2-D applications, to rectangular signals of possibly non-dyadic sides.
 Typically, the standard convolution-based NDWT can only manage dyadic or squared 2-D  input signals of dyadic scale (e.g., Wavelab).

\section{Two Examples of Application}
\label{sec:twoexamples}
In this section, we provide two applications in which the package {\bf WavmatND} is used. In the first application we apply our matrix-based NDWT to obtain a scaling index from the background of a mammogram image. The scaling index of an image is measured by Hurst exponent, a dimensionless constant in interval
$[0,1].$ For locally isotropic medical images, the Hurst exponent is known to be useful for diagnostic purposes (\cite{ramirez2007wavelet},\cite{nicolis2011}, \cite{jeon2014mammogram}). Wavelet-based spectra of an image is defined on a selected hierarchy of multiresolution spaces
in a wavelet representation as
a set of pairs $(j, S(j))$, where $j$ is the multiresolution
level and $S(j)$ is the logarithm of the average of squared wavelet coefficients at that level.
The Hurst exponent, as a measure of regularity of the image, is functionally connected with the slope
of a linear fit on pairs $(j, S(j))$.
Any type of wavelet decomposition can serve as a generator of wavelet spectra, and in this application we look at 2-D scale-mixing NDWT of a digital mammogram.

\begin{figure}[h]
\centering
\includegraphics [width=13cm, height=10 cm]{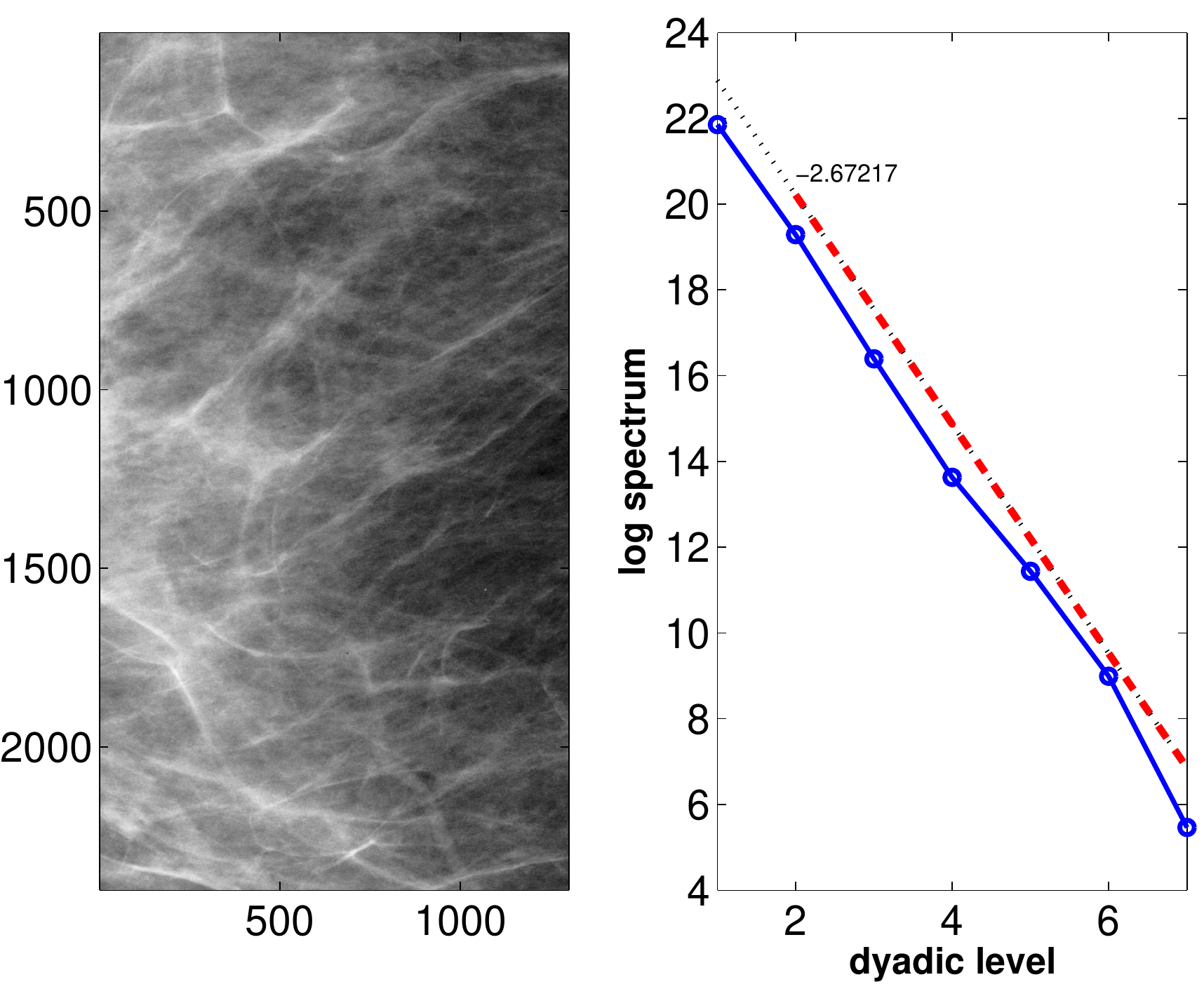}
\caption{Left panel: The selected region of interest (ROI) where the Hurst exponent is estimated. Right panel: The dash-dotted line represents 2-D non-decimated wavelet spectra of the ROI from the left panel. The dashed line shows the regression result using the corresponding energy levels.} \label{spectra}
\end{figure}

The digital mammogram analyzed comes from the Digital Database for Screening Mammography (DDSM) at the University of South Florida.
  The image is digitized  by HOWTEK scanner at the full 43.5-micron per pixel spatial resolution and features craniocaudal (CC) projection. A detailed description of the data can be found in \cite{Heath2000}.
\begin{figure}[h]
\centering
\includegraphics [width=4.5cm, height=9 cm]{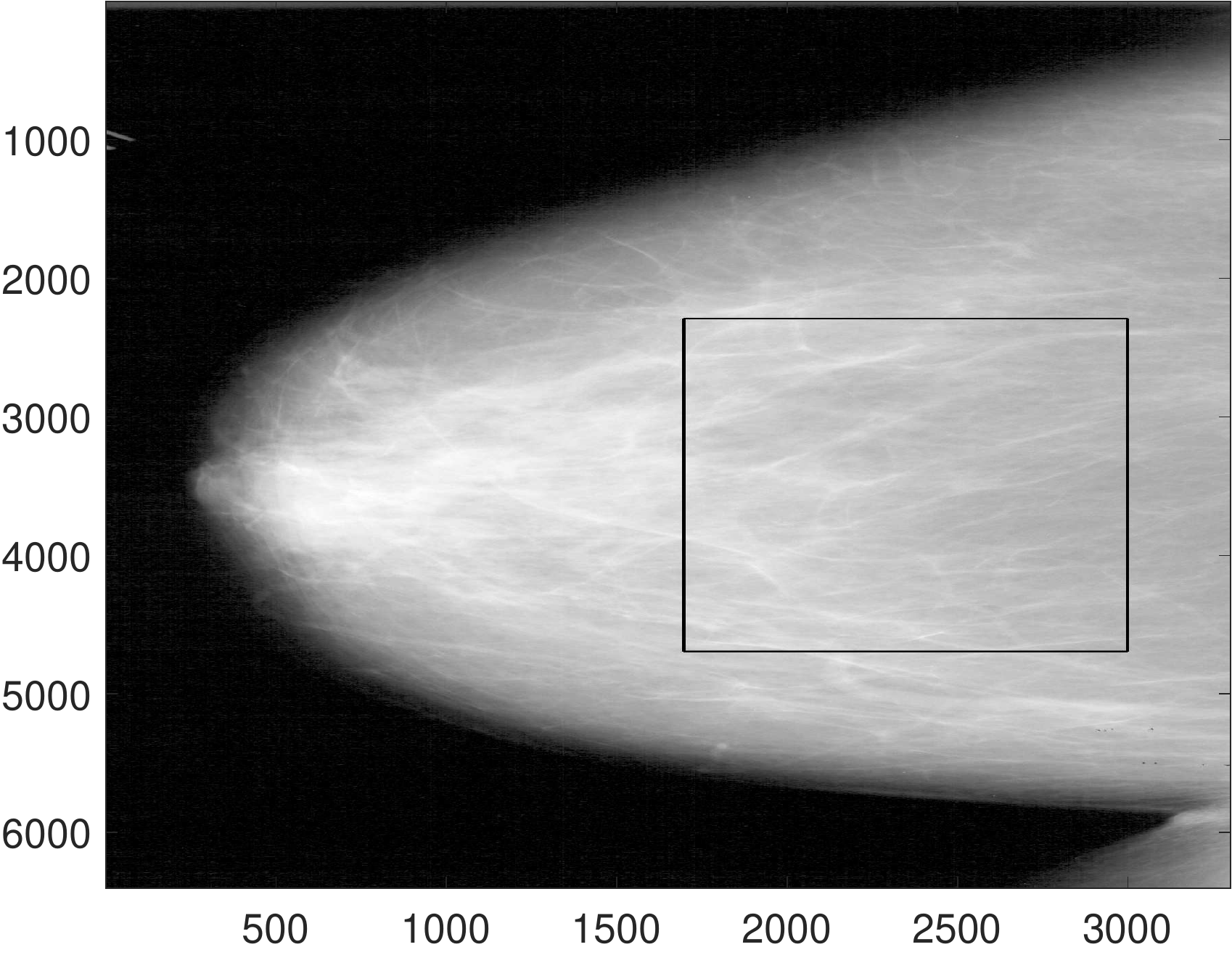}
\caption{A ROI in a mammogram image selected for the estimation of scaling.} \label{tissue}
\end{figure}
Figure~\ref{tissue} shows the location of the region of interest (ROI) within the mammogram. We selected the ROI of size $2401 \times 1301$ and transformed it to a scale-mixing 2-D non-decimated wavelet domain.
 The spectral slope is estimated from the levelwise log-average squared
 coefficients along the diagonal hierarchy of multiresolution spaces, comprising the wavelet spectra, as in Figure~\ref{spectra}.
 The slope of $-2.6722$ gives the Hurst exponent of $-(slope+2)/2 = 0.3361$. Details can be found
  in \cite{minkyoung:thesis} who use the Hurst exponent estimators to classify the mammograms from the DDSM data base for breast cancer detection.

In the second example, we denoise a signal captured by an atomic force microscope. The atomic force microscopy (AFM) is a type of scanned proximity probe microscopy that  measures the adhesion strength between two materials at the nanonewton scale.
  The AFM data from the adhesion measurements between carbohydrate and the cell adhesion molecule (CAM) E-Selectin was collected by Bryan Marshall from the   Department of Biomedical Engineering at Georgia Institute of Technology. The technical description and details are provided in \cite{marshall2005force}.

 \begin{figure}[h]
\centering
\includegraphics [width=10.2cm, height=9.3 cm]{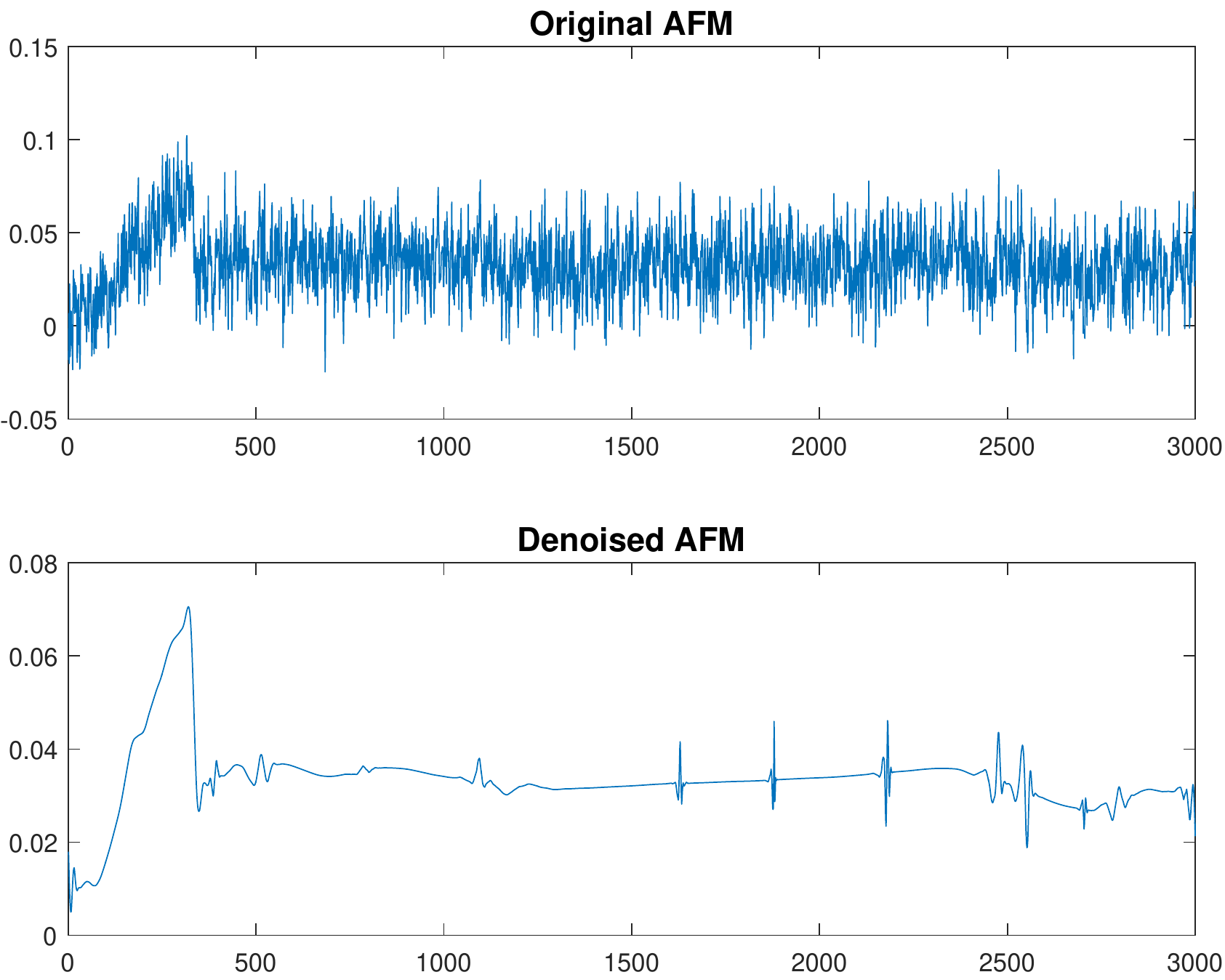}
\caption{Denoising of AFM by hard-thresholding on NDWT coefficients with 6-tab Daubechies wavelet. } \label{fig:afm}
\end{figure}

 In AFM, a cantilever beam is adjusted until it bonds with the surface of a sample, and then, the force required to separate the beam and sample is measured from the beam deflection. Beam vibration can be caused by external factors such as thermal energy of the surrounding air and the footsteps of someone outside the laboratory. The vibration of a beam shows as noise on the deflection signal.  For denoising purposes, we decomposed AFM signal of size 3,000 into $10$ decomposition levels using the NDWT with a 6-tab Daubechies wavelet (3 vanishing moments) and applied hard thresholding on wavelet coefficients. The threshold for this process is set as $\sqrt{2\log m}\hat{\sigma}$, where $\hat{\sigma}$ is an estimator of standard deviation of noise present in the wavelet coefficients at the finest level of detail, and $m$ is the size of the original signal.
 Given the redundancy of the transform, we estimate $\hat{\sigma}$ by averaging two estimators,
  $\hat{\sigma}_o$ and $\hat{\sigma}_e$, which are sample standard deviations of wavelet coefficients at every odd and even locations, respectively, within the finest level of detail.
  Figure \ref{fig:afm} shows the noisy AFM signal and its denoised version. The researchers are particularly interested in
  the shape of the signal for the first 350 observations of an AFM signal, prior to cantilever detachment.

  \section{Package Description and Demos}
  The {MATLAB} package, {\bf WavmatND},  includes two core functions, several
  additional functions, and data sets needed for illustrative examples and demos.
  \vspace{0.1in}  \\
  \subsection{Core Functions}
  \texttt{WavmatND()} is a core function that generates a transform matrix.  Inputs to this function are a wavelet filter, size of an input signal, the depth of transformation, and a shift. The shift
  corresponds to parameter $s$ in the definition of quadrature mirror filter $g_i = (-1)^{l-i} h_{M-s-i},~
  $ $ i=0, \dots, M,$  and is usually taken as 0 or 2.
  \vspace{0.1in} \\
  \texttt{weight()} generates a weight matrix that rescales every submatrix  in the inverse wavelet transform. This matrix is necessary for the lossless inverse transform, as in (\ref{eq:2d}). It assigns different weights to each submatrix, as described in Section \ref{sec:mf2d}. Inputs to this function are the size of the original signal and the depth of the transform.\vspace{0.1in} \\
  \subsection{Other Functions and Data Sets Included}

  For the illustration purposes, we include a custom made function {\tt WaveletSpectra2NDM.m} for assessing the scaling in images based on 2-D NDWT.

  \texttt{WaveletSpectra2NDM()} estimates a scaling index of an image using the diagonal hieararchy of nested multiresolution spaces in a 2-D scale-mixing NDWT. It returns the average level-energies for a specified range of levels, scaling slope, and a graph showing linear regression fit of log energies on the selected levels.  The inputs are 2-D data/image, the depth of transform, a wavelet filter, a range of levels used for the regression, and an option for showing the plot (1 for a plot and 0 for no plot). \vspace{0.1in} \\
  \texttt{NDWT2D()} is a function that performs a standard 2-D NDWT using NDWT matrices. It returns $c$-, $h$-, $v$-, and $d$-types of wavelet coefficients. In this transform there is no scale mixing and $x$-scale is the same as the $y$-scale. Inputs to this function are an image, a wavelet filter, the depth of transform, and a shift.\vspace{0.1in} \\
  {\tt filters.m} contains some commonly used wavelet filters needed for construction of a NDWT matrix. It provides
  Haar,
   Daubechies 4-20, Symmlet 8-20, and Coiflet 6, 12, and 18 filters with high accuracy.
      Users can choose an appropriate wavelet filter based on type of analysis and input data, compromising between the smoothness and locality. \vspace{0.1in} \\
  {\tt afm.mat} and {\tt tissue.mat} are data sets used in the two applications. Interested readers can load the data sets for further analysis. We also included the code used to generate the results in the paper at {\tt exampleApplications.m}. \vspace{0.1in} \\
  {\tt lena.mat} is well-known image of Lena S\"oderberg, one of the most used images in signal processing community.
  This image is utilized in DEMO 1 explained in the next section.



  \subsection{DEMO 1: Transform and reconstruction}
\begin{figure}[h]
\centering
\includegraphics [width=14cm,height=8cm , clip]{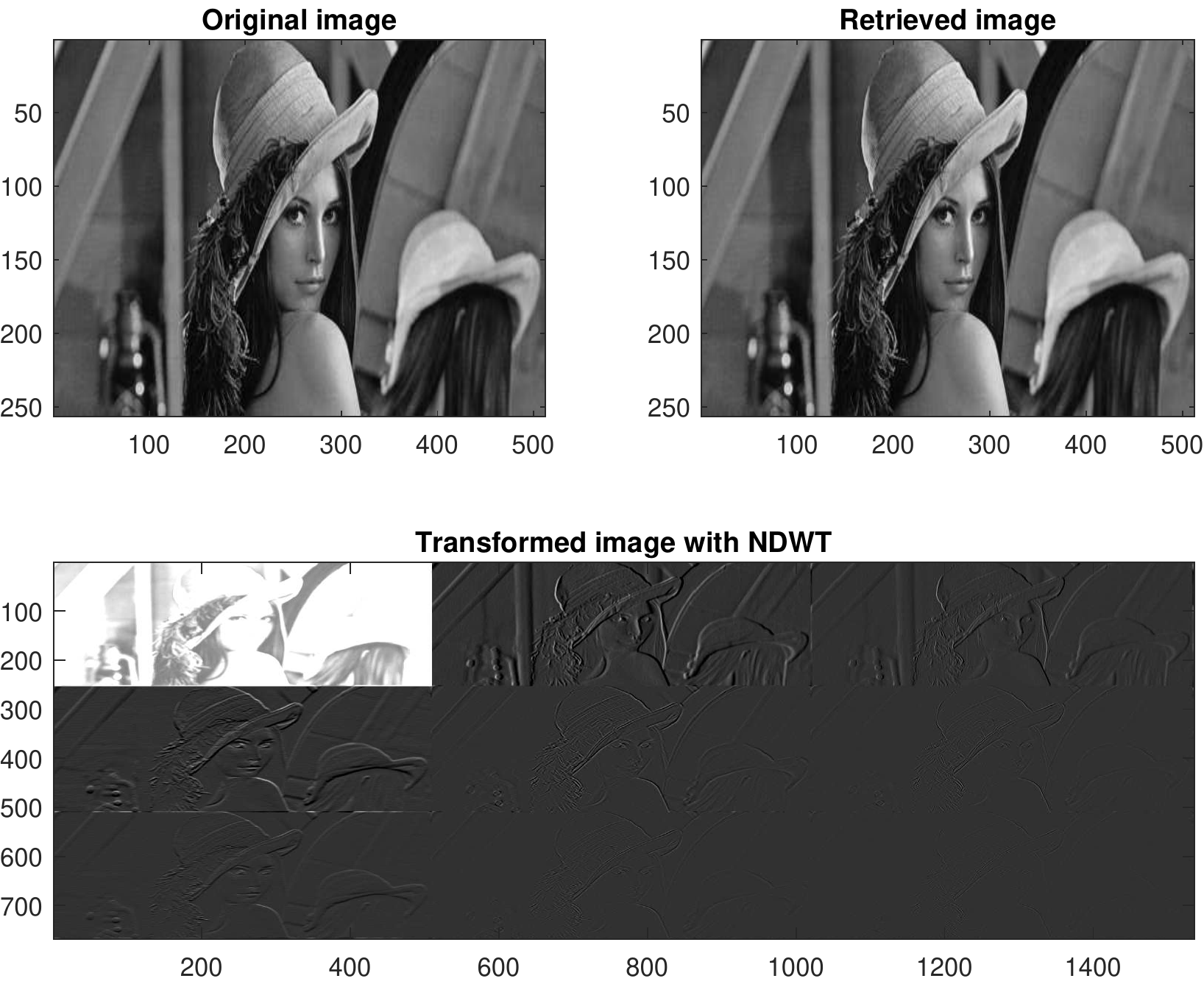}
\caption{Lena image in the original and wavelet domains from Demo 1.}
\label{fig_demo1}
\end{figure}

  As we discussed earlier, a matrix-based NDWT maps an original data set into a time-scale domain with efficient and simple steps. In the following code, we load image \texttt{lena}, of size $(256 \times 512)$ and create two NDWT matrices \texttt{W1} and \texttt{W2} that perform the NDWT on image by columns and rows, respectively. We use the Haar wavelet and perform a $p$-depth NDWT in both columns and rows for $p=\log(\min(m,n))-1=7$.\\
  \texttt{load lena;   [n m]=size(lena);\\
  p=floor(log(min(m,n)))-2; shift=0; \\
  h = [1/sqrt(2) 1/sqrt(2)];\\
  W1=WavmatND(h,n,p,shift); W2=WavmatND(h,m,p,shift); \\
  tlena=W1*lena*W2'; }   \\
  The reconstruction of the transformed lena \texttt{tlena} is simple. We generate weight matrices, \texttt{T1} and \texttt{T2}, of the sizes compactible with \texttt{W1} and \texttt{W2}, respectively, and reconstruct the signal sa follows: \\
  \texttt{T1=weight(n,p);	T2=weight(m,p);\\
  rlena=W1'*T1*tlena*T2*W2; }\\
  The reconstructed signal is \texttt{rlena}. The transformation and reconstruction are illustrated in Figure \ref{fig_demo1}.

  \subsection{DEMO 2: Denoising of Doppler Signal}
  In this demo, we first generate a Doppler signal \texttt{s} of size 1,000 and a matching signal \texttt{noise}
  consisting of i.i.d normal variables with mean 0 and variance $0.05^2.$
  The sum of \texttt{s} and \texttt{noise}
  constitutes a noisy signal \texttt{sn} with signal-to-noise ratio of 5.78. \\
  \texttt{sigma=0.05; m=250; \\
  t = linspace(1/m,1,m);\\
  s = sqrt(t.*(1-t)).*sin((2*pi*1.05) ./(t+.05));\\
  noise=normrnd(0,1,size(s))*sigma;\\
  sn=s+noise; }\\
  Next, with a Haar wavelet, we generate the NDWT matrix, \texttt{W}, which decomposes the signal into $\big(\lfloor \log (1,000) \rfloor -1 \big)$ decomposition levels. The resulting wavelet coefficients are in \texttt{tsn}. \\
  \texttt{J=floor(log2(m)); k=J-1;\\
  qmf = [1/sqrt(2) 1/sqrt(2)]; \\
  W = WavmatND(qmf,m,k,0);  \\
  T = weight(m, k);\\
  tsn=W*sn'; }\\
  Then, we apply hard thresholding for denoising. Hard thresholding is applied to all detail level subspaces, and the threshold is set to be $\sqrt{2\log(m)}\hat{\sigma}$, where $\hat{\sigma}$ is the square root of average of variances of wavelet coefficients at  odd  and even positions at the finest level of detail, and $m$ is the length of the original signal.\\
  \texttt{sigma2hat=(var(tsn(end-m+1:2:end))+var(tsn(end-m:2:end)))/2;\\
  threshold=sqrt(2*log(k*m)*sigma2hat); \\
  snt= tsn(m+1:end).*(abs(tsn(m+1:end))>threshold); \\
  rs=W'*T*[tsn(1:m); snt];} \\
  The reconstructed denoised signal is \texttt{rs}.

\begin{figure}[h]
\centering
\includegraphics [width=14.5cm,height=8cm , clip]{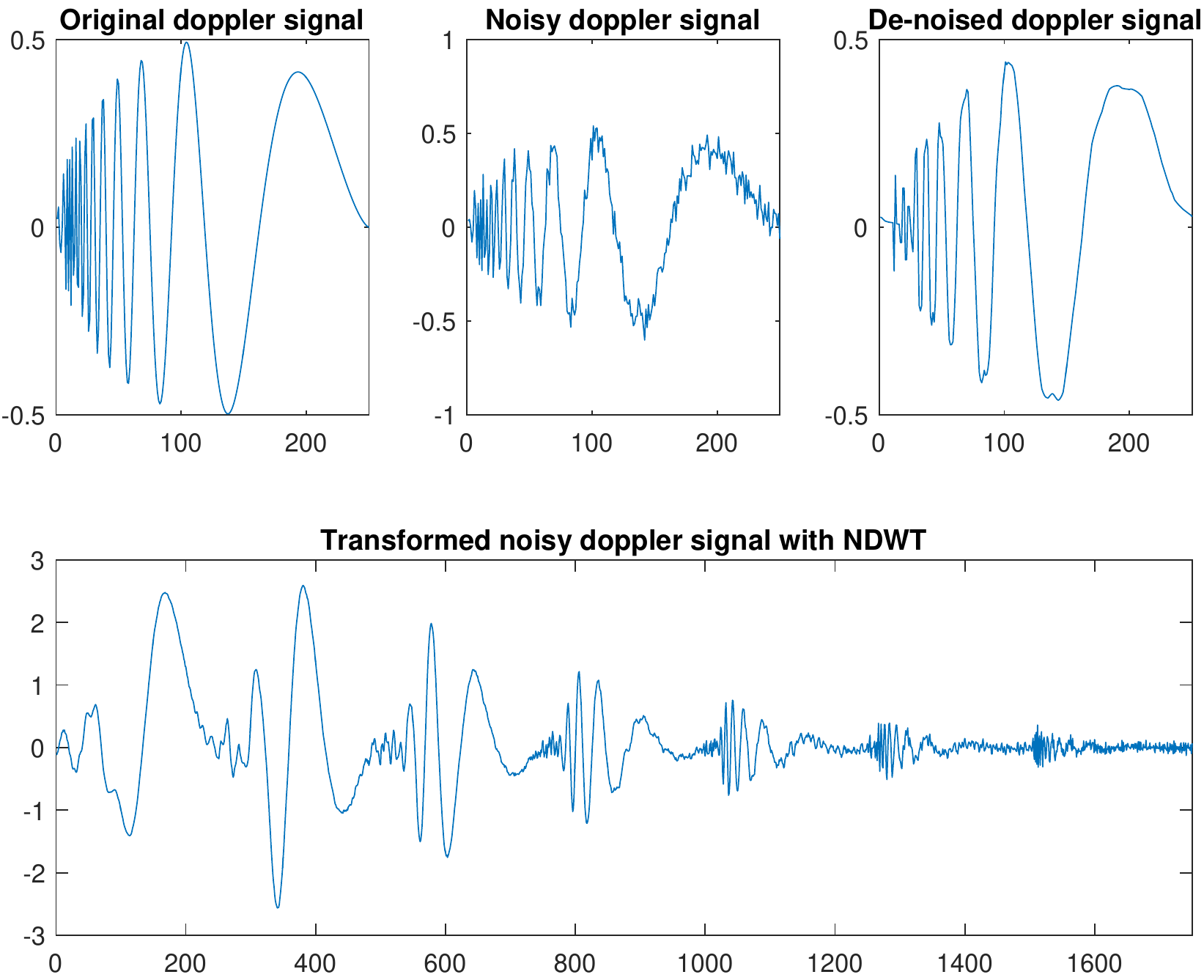}
\caption{ Doppler signals in time and wavelet domains for denoising from Demo 2.} \label{fig_demo2}
\end{figure}
%

\section{Discussion}
The non-decimated wavelet transform (NDWT) possesses properties beneficial in various wavelet applications. We developed   {MATLAB} package,{\bf WavmatND}, which performs the NDWT in one or two dimensions. Instead of repeated convolutions that are standardly performed, the NDWT is performed by matrix multiplication. This significantly decreased the computation time in simulations
when performed in {MATLAB} computing environment. This reduction in computation time is additionally augmented when we applied the NDWT repeatedly to signals of the same size, decomposition level, and choice of wavelet basis.
 In 2-D case, the NDWT matrix yields a scale-mixing NDWT, which turns out to be more compressive compared to the standard
 2-D NDWT. For lossless retrieval of an original signal, we utilize
 a weight matrix.  We also relax the constraint on the size of input signals so that the NDWT could be performed on signals of non-dyadic  size in one or two dimensions.  We hope that this stand-alone {MATLAB} package will be a useful tool for practitioners interested in various aspects of signal and image processing.

 The package {\bf WavmatND} can be downloaded from Jacket's Wavelets WWW repository site   \url{http://gtwavelet.bme.gatech.edu/}.

\bibliography{WM}

\end{document}